\documentclass[a4paper,12pt,onecolumn,final,notitlepage,oneside]{article}
\usepackage[english]{babel}
\usepackage[cp1251]{inputenc}
\usepackage[T2A]{fontenc}
\usepackage{graphicx}
\usepackage[hidelinks]{hyperref}
\usepackage{amsmath}%
\usepackage{amsfonts}%
\usepackage{amssymb}%
\begin{document}
\title{Primordial Magnetic Field Generated in Natural Inflation}
\author{{Anwar S AlMuhammad \thanks {anwar@physics.utexas.edu}} , {{Rafael Lopez-Mobilia\thanks{Rafael.LopezMobilia@utsa.edu}}}}
\maketitle
\begin{center}
Department of Physics and Astronomy, The University of Texas at San Antonio (UTSA), One UTSA Circle, San Antonio, TX 78249, USA

Received: June 9, 2015. Accepted: October 2, 2015

Published in the Journal General Relativity and Gravitation, Volume 47, Issue 11  
\end{center}

\begin{abstract}
We study the simple gauge invariant model ${f^2}FF$ as a way to generate primordial magnetic fields (PMF) in Natural Inflation (NI). We compute both magnetic and electric spectra generated by the ${f^2}FF$ model in NI for different values of model parameters and find that both de Sitter and power law expansion lead to the same results at sufficiently large number of e-foldings. We also find that the necessary scale invariance property of the PMF cannot be obtained in NI in first order of slow roll limits under the constraint of inflationary potential, $V\left( 0 \right) \simeq 0$. Furthermore, if this constraint is relaxed to achieve scale invariance, then the model suffers from the backreaction problem for the co-moving wave number, $k \lesssim 8.0\times 10^{-7} \rm{Mpc^{-1}}$ and Hubble parameter, $H_i \gtrsim 1.25\times 10^{-3} \rm{M_{\rm{Pl}}}$. The former can be considered as a lower bound of $k$ and the later as an upper bound of $H_i$ for a model which is free from the backreaction problem. Further, we show that there is a narrow range of the height of the potential $\Lambda $ around ${\Lambda _{\min }} \approx 0.00874{M_{{\rm{Pl}}}}$ and of $k$ around ${k_{\min }} \sim 0.0173{\rm{Mp}}{{\rm{c}}^{ - 1}}$, at which the energy of the electric field  can fall below the energy of the magnetic field. The range of $k$ lies within some observable scales. However, the relatively short range of $k$ presents a challenge to the viability of this model.
\end{abstract}


PACS: 

\section{Introduction}

Inflationary cosmology has solved many of the fundamental problems of the Big Bang model \cite{1}. Inflationary theory provides an explanation for the large scale structure of the Universe, which is linked to the quantum fluctuation of the field of inflation $\phi $. The nearly flat, homogeneous, and isotropic universe, and the scale invariant spectrum of the anisotropies of the CMB, as well as the generation of primordial gravitational waves (PGW), are all predictions of inflation. All of the above predictions (with the possible exception of PGW) have been confirmed by various cosmological observations. 

Similarly, magnetic fields are being observed in all kinds of galaxies and cluster of galaxies at wide range of redshifts. Moreover, a lower bound, $B \ge 3 \times {10^{ - 16}}{\rm{ G}}$, has been reported for intergalactic magnetic fields \cite{2}. Planck, 2015 results constrains the upper limit to be of the order of $B < {10^{ - 9}}{\rm{ G}}$ \cite{2}. The commonly accepted model for the generation of galactic magnetic fields is the galactic dynamo \cite{3}. This model needs a seed field to start amplifying the magnetic field. Also, it cannot explain the existence of fields in the absence of a uniform rotating charged medium, like the magnetic fields indicated in voids. Hence, such a detection may indicate a cosmological source of primordial origin. The generation of such a large scale primordial magnetic fields (PMF) is still an open question, see the reviews of this subject \cite{4}. However, one of the most interesting cosmological models is the simple ${f^2}FF$ model \cite{5}-\cite{7}. It has gained more interest because it is a stable model under perturbation. (See \cite{8} and references therein.) Also, it can lead to a scale invariant spectrum of PMF under the exponential potential model of inflation \cite{7}. 

The main difficulties with this model are the backreaction problem, where the scale of the associated electric field with PMF spectrum can exceed the scale of inflation itself \cite{7,9}, and the strong coupling between electromagnetic fields and charged matter at the beginning of inflation \cite{10}. In this case, if the electromagnetic field couples to charged matter, the physical charge associated with it is huge at the onset of inflation. For example, for the number of e-folds of inflation, $N = 60$, the physical electric charges, $q \propto {e^{120}}$ \cite{10}. Such an absurdly large charge makes this model not viable.

The starting point is the Lagrangian of a scalar (inflaton) field $\phi $ coupled to the electromagnetic (vector) field ${A_\mu }$ \cite{7} which can be written as,

\begin{equation}
{\cal L} =  - \sqrt { - g} \,\ \Big{[}  \frac{1}{2}\left( {{\partial _\mu }\phi } \right)\left( {{\partial ^\mu }\phi } \right) + V\left( \phi  \right) + \frac{1}{4}{g^{\alpha \beta }}{g^{\mu \nu }}{f^2}\left( {\phi ,{\rm{ }}t} \right){F_{\mu \alpha }}{F_{\nu \beta }} \Big{]},
\label{eqn1}
\end{equation}
where, ${F_{\nu \beta }} = {\partial _\nu }{A_\beta } - {\partial _\beta }{A_\nu }$ is the electromagnetic field tensor and $g$ is the determinant of the spacetime metric ${g_{\mu \nu }}$. The first term in the Lagrangian is the standard kinetic part of the scalar field, and the second term,$V\left( \phi  \right)$, is the potential. A Lagrangian of a pure electromagnetic field would be of the form $ - \frac{1}{4}{F_{\mu \nu }}{F^{\mu \nu }}$, but here we couple it to the scalar field through the unspecified function $f(\phi ,{\rm{ }}t)$. The main reason behind this coupling is to break the conformal symmetry of electromagnetism and hence prevent the dilution of the seed of magnetic field as it is generated in the inflation era.  

Simple models of inflation potential are based on a single scalar field, such as quadratic, $V\left( \phi \right) \sim {\phi^2}$, quartic, $V\left( \phi  \right)\sim{\phi^4}$ \cite{11,13}, Higgs potential, $V\left( \phi  \right)\sim M^4\left(1-\exp{[-\sqrt{2/3} \phi/M_{PL}]}\right)^{2} $ \cite{12} and the exponential potential, $V\left( \phi  \right)\sim\exp [ - \sqrt {2{\varepsilon _1}} \left( {\phi  - {\phi _0}} \right)]$ \cite{13}. The last one is used in \cite{6,7} to find the magnetic and electric spectrum in the ${f^2}FF$ model. These models became more interesting after WMAP9 \cite{14} and Planck \cite{15}. As a result, the preferred potential class is the so called "\textit{plateau inflation}", at which $V\left( 0 \right) \ne 0$. 

In March 2014, results from the Background Imaging of Cosmic Extragalactic Polarization (BICEP2) experiment were released \cite{16}. They reported the detection of the tensor mode (B-mode) polarization of temperature anisotropies in the CMB. The tensor to scalar ratio reported by BICEP2 was $r = 0.2_{ - 0.05}^{ + 0.07}$, with $r = 0$ disfavored at $7.0\sigma $. Also, the scale of inflation energy is close to the Grand Unified Theory GUT scale, ${\rho _{{\rm{GUT}}}}^{1/4} \sim 10^{16}\rm{GeV}$. The results of BICEP2 put many standard inflationary models in trouble \cite{17}. As a result, non-standard models, such as Large Field Inflation (LFI) \cite{11, 18}, Natural Inflation (NI) \cite{19} models are a better fit with the BICEP2 results.

On the other hand, Planck results released in Sep 2014 \cite{20} cast serious doubts on the primordial origin of this polarization signal. These results of Planck indicate that there is a significant contamination by dust over most of the high Galactic latitude sky in the same region where BICEP2 detected B-mode polarization. Consequently, there is a good chance that the source of the observations reported by BICEP2 is all galactic dust and not of primordial origin.

Finally, the joint analysis of BICEP2/Keck Array and Planck (BK/P) data was released on Feb 2015 \cite{21}. The joint data of three probes eliminate the effect of dust contamination and show that the upper limit of tensor to scalar ratio, ${r_{0.05}} < 0.12$ at 95\% CL, and the gravitational lensing B-modes (not the primordial tensor) are detected in $7\sigma $[21]. Similarly, the inflationary models are constrained based on the new analysis of data. The scalar spectral index was constrained by Planck, 2015 to be ${n_s} = 0.9682 \pm 0.0062$\cite{22}. Consequently, the more standard inflationary models, like ${R^2}$-inflation, which result low value of $r$, are the most favored one by Planck, 2015. However, the chaotic inflationary models like LFI and NI are disfavored \cite{22}. These results ruled out the first results of BICEP2, 2014. 

In this paper, the simple inflation model, ${f^2}FF$, of PMF will be investigated in detail under the natural inflation (NI), in the same way as done in \cite{7} and as we did in the context of LFI \cite{23} and ${R^2}$-inflation \cite{24}. We adopt natural units, $\left[ {c = \hbar  = {k_B} = \;1} \right]$, the signature $( - 1,{\rm{ 1, 1, 1)}}$, and a spatially-flat universe, where the reduced Planck mass, ${M_{{\rm{Pl}}}}{\rm{\;}} = {\left( {8\pi G} \right)^{ - 1/2}}$,will be taken as unity in the computations. Hence, the potential of NI can be written \cite{12, 19} as,
\begin{equation}
V\left( \phi  \right) = {\Lambda ^4}[ {1 + \cos (\frac{\phi }{\sigma })} ],
\label{eqn2}
\end{equation}
where $\Lambda$ is fixed by CMB normalization of the potential, and $\sigma$ is the mass scale of the model. If $\sigma$ is taken to be of the order of $M_{\rm{Pl}}$, then $\Lambda \sim 10^{-13} M_{\rm{Pl}}$. Since Eq.(\ref{eqn2}) is a periodic, even function of $\phi $, one can study the potential in the interval $\phi  \in \left[ {0,\pi \sigma } \right]$ \cite{12}. In order to reach the GUT scale of inflation (${10^{15}} - {10^{16}}{\rm{GeV}}$), the mass scale has to be of the order of $\sigma \sim {M_{{\rm{Pl}}}}$ \cite{19}. 

The order of the paper will be as follows. In section \ref{sec:slow roll}, the slow roll natural inflation formulation will be presented for both the simple de Sitter model of expansion and the more general, power law expansion. In section \ref{sec:PMF in NI}, the PMF and associated electric fields are computed for NI at different values of the parameters. In section \ref{sec:Summary}, we summarize and discuss the results.

\section{Slow roll analysis of Natural Inflation}
\label{sec:slow roll}

During inflation, we will assume the electromagnetic field to be negligible compared to the scalar field, $\phi$ \cite{7}. Hence, the equation of motion derived from (\ref{eqn1}) for the scalar field can be written as,
\begin{equation}
\ddot \phi  + 3H\dot \phi  + {V_\phi } = 0,
\label{eqn3}
\end{equation}
where, $H\left( t \right) = \;\dot a\left( t \right)/a\left( t \right)$, is the Hubble parameter as a function of cosmic time, $t$, and $a\left( t \right)$ is the cosmological scale factor. The over dot indicates differentiation respect to cosmic time, and ${V_\phi } = {\partial _\phi }V$. The Friedman equation can be obtained from the Einstein field equations by assuming a spatially-flat Friedmann-Robertson-Walker FRW universe, which yields,
\begin{equation}
H{\;^2} = \;\frac{1}{{3{M_{{\rm{Pl}}}}^2}}\left[ {\frac{1}{2}{{\dot \phi }^2} + V\left( \phi  \right)} \right].
\label{eqn4}
\end{equation}
During inflation, the slow roll approximation allows us to neglect the second derivative in (\ref{eqn3}), which leads to the attractor condition
\begin{equation}
\dot \phi  \approx  - \frac{{{V_\phi }}}{{3H}}.
\label{eqn5}
\end{equation}

Defining the slow roll parameters of inflation in terms of the potential \cite{12,25}, of a single inflation field for NI (\ref{eqn2}),
\begin{equation}
{\epsilon _{1V}}\left( \phi  \right) = \frac{1}{2}{M_{{\rm{Pl}}}}^2{\left( {\frac{{{V_\phi }}}{V}} \right)^2} = \frac{1}{{2{\zeta ^2}}}{\left( {\frac{{\sin (\frac{\phi }{\sigma })}}{{1 + \cos (\frac{\phi }{\sigma })}}} \right)^2},
\label{eqn6}
\end{equation}
\begin{equation}
{\epsilon _{2V}}\left( \phi  \right) = 2{M_{{\rm{Pl}}}}^2\left( {{{\left( {\frac{{{V_\phi }}}{V}} \right)}^2} - \frac{{{V_{\phi \phi }}}}{V}} \right) = \frac{2}{{{\zeta ^2}}}\frac{1}{{\cos (\frac{\phi }{\sigma }) + 1}},
\label{eqn7}
\end{equation}
where $\zeta  \equiv \sigma /{M_{{\rm{Pl}}}}$. These slow roll parameters can be written in terms of the Hubble parameter,
\begin{equation}
{\epsilon _{1H}}\left( \phi  \right) = 2{M_{{\rm{Pl}}}}^2{\left( {\frac{{{H_\phi }}}{H}} \right)^2},{\rm{    }}{\epsilon _{2H}}\left( \phi  \right) = 2{M_{{\rm{Pl}}}}^2\left( {\frac{{{H_{\phi \phi }}}}{H}} \right),
\label{eqn8}
\end{equation}
The relation between the two formalisms \cite{25} can be written as
\begin{equation}
{\epsilon _{1V}} = {\epsilon _{1H}}{\left( {\frac{{3 - {\epsilon _{2H}}}}{{3 - {\epsilon _{1H}}}}} \right)^2}.
\label{eqn9}
\end{equation}

All of the above parameters are assumed to be very small during the slow roll inflation,$({\epsilon _{1V}},{\rm{ }}{\epsilon _{2V}},{\rm{ }}{\epsilon _{1H}}{\rm{, }}{\epsilon _{2H}}) \ll 1$. Further, inflation ends when the values of $\left\{ {{\epsilon _{1V}}{\rm{ }},{\rm{ }}{\epsilon _{1H}}} \right\} \to 1$. At first order one can neglect ${\epsilon _{1H}}$ and ${\epsilon _{2H}}$as compared with 3, obtaining ${\epsilon _{1V}} \approx {\epsilon _{1H}}$. Therefore, using (\ref{eqn8}) and the relation between the cosmic time, $t$, and the conformal time,$\eta $, $dt = a\left( \eta  \right)d\eta $, one can write the relation between conformal time and slow roll parameter, ${\epsilon _{1H}}$, as \cite{18},
\begin{equation}
\eta  =  - \frac{1}{{aH}} + \int {\frac{{{\epsilon _{1H}}}}{{{a^2}H}}da}.
\label{eqn10}
\end{equation}
Assuming, ${\epsilon _{1H}} \approx const$, and then integrating (\ref{eqn10}) yields the power law expansion of the universe during inflation,
\begin{equation}
a(\eta ) = {l_0}{\left| \eta  \right|^{ - 1 - {\epsilon _{1H}}}},
\label{eqn11}
\end{equation}
where, ${l_0}$ is the integration constant. 

On the other hand, in the simplest form of inflationary expansion (de Sitter), the universe expands exponentially during inflation at a very high but constant rate, ${H_i}$, 
\begin{equation}
{H_i} = \frac{{\dot a}}{a} \approx {\rm{const}},
\label{eqn12}
\end{equation}
\begin{equation}
a\left( t \right) = a\left( {{t_1}} \right)\;{\rm{exp}}\left[ {{H_i}t} \right],
\label{eqn13}
\end{equation}
where ${t_1}$ is the starting time of inflation. In conformal time, Eq.(13) can be written as,
\begin{equation}
a\left( \eta  \right) =  - \frac{1}{{{H_i}\;\eta \;}}.
\label{eqn14}
\end{equation}
Plugging (14) into the relation between cosmic and conformal time and integrating implies that $\eta  \to \left( { - \infty ,{\rm{ }}{0^ - }} \right)$ as $t \to \left( {0,{\rm{ }}\infty } \right)$. 
	
As inflation ends when ${\epsilon _{1V}}\left( \phi  \right) \approx 1$, then from Eq.(\ref{eqn6}),
\begin{equation}
{\phi _{end}} \approx \sigma {\rm{ arccos}}\left( {\frac{{1 - 2{\zeta ^2}}}{{1 + 2{\zeta ^2}}}} \right).
\label{eqn15}
\end{equation}
If ${\phi _{end}} \ll 1,$ then ${\sigma _{end}} \ll 1$, or ${\sigma _{end}} \to \frac{{{M_{{\rm{Pl}}}}}}{2}\left( {1 - \frac{{(2n + 1)}}{2}\pi } \right)$, where, $n =$integer. Also, the relations between the slow roll parameters and the scalar spectral index,${n_s}$, and tensor-to-scalar ratio, $r$, can be written as follows \cite{25},
\begin{equation}
{n_s} = 1 - 6{\epsilon _{1V}} + 2{\epsilon _{2V}},
\label{eqn16}
\end{equation}
\begin{equation}
r = 16{\epsilon _{1V}}.
\label{eqn17}
\end{equation}
Substitution of (\ref{eqn6})-(\ref{eqn7}) into (\ref{eqn16})-(\ref{eqn17}) yields
\begin{equation}
{n_s} = 1 + \frac{3}{{{\zeta ^2}}} - \frac{1}{{{\zeta ^2}}}{\sec ^2}\left( {\frac{\phi }{{2\sigma }}} \right),
\label{eqn18}
\end{equation}
\begin{equation}
r = \frac{8}{{{\zeta ^2}}}{\left( {\frac{{\sin (\frac{\phi }{\sigma })}}{{1 + \cos (\frac{\phi }{\sigma })}}} \right)^2}.
\label{eqn19}
\end{equation}

One can find the relation between $r$ and ${n_s}$ which depends on the number of e-folds of inflation, $N$. The first order of approximation for $N$ can be written \cite{12,25} as, 
\begin{equation}
N \approx  - \sqrt {\frac{1}{{2{M_{{\rm{Pl}}}}^2}}} \;\mathop \smallint \limits_\phi ^{{\phi _{end}}} \frac{1}{{\sqrt {{\epsilon _1}} }}d\phi \; = {\zeta ^2}\ln \left[ {\frac{{1 - \cos (\frac{{{\phi _{end}}}}{\sigma })}}{{1 - \cos (\frac{\phi }{\sigma })}}} \right],
\label{eqn20}
\end{equation} 
where $N$ is the difference between the final e-fold and the e-fold at $t$. Solving for $\phi $ in (\ref{eqn20}) and substituting ${\phi _{end}}$ from (\ref{eqn15}),
\begin{equation}
\phi  = \sigma {\rm{ arc}}cos\left( {1 - [1 - \cos (\frac{{{\phi _{end}}}}{\sigma })]\exp ( - N/{\zeta ^2})} \right).
\label{eqn21}
\end{equation}
Now, combine (\ref{eqn18}) and (\ref{eqn19}), 
\begin{equation}
r = \frac{8}{3}\left\{ {{n_s} - 1 + \frac{1}{{{\zeta ^2}}}{{\sec }^2}\left( {\frac{\phi }{{2\sigma }}} \right)} \right\}{\left(\frac{\sin (\frac{\phi }{\sigma })}{\left\{ {1 + \cos (\frac{\phi }{\sigma })} \right\}}  \right)^2}.
\label{eqn22}
\end{equation}
By substituting (\ref{eqn21}) into (\ref{eqn22}), one can plot $r - {n_s}$ for different values of  $\zeta$ and $N$.

Shortly after the onset of inflation the value of $H$ becomes very high and is approximately constant, but later on it decreases as the value of the field changes. Also, after few ${N_ * }$, the spacetime (pivot scale, ${k_ * }$) exits from the Hubble horizon. We adopt Planck, 2015 pivot scale, ${k_ * } = 0.05{\rm{Mp}}{{\rm{c}}^{ - 1}}$. One can consider $H$ as a constant, after the first few e-foldings. That is basically the de Sitter expansion, which is exactly exponential expansion as described by (\ref{eqn13}). So, it is worthwhile to investigate both cases; the de Sitter, and the more realistic power law model described by (\ref{eqn11}).

\subsection{NI under de Sitter expansion}
\label{subsec:sr in dS}

The de Sitter model is the zeroth order approximation, which does not have graceful exit from inflation \cite{26}. But it can be used as an approximation at the early stages of inflation. In conformal time, $\eta $, Eq.(\ref{eqn5}) can be written as,
\begin{equation}
\frac{1}{{a\left( \eta  \right)}}\phi ' \approx  - \frac{{{V_\phi }}}{{3{H_i}}},
\label{eqn23}
\end{equation}
where, $\phi ' = {\partial _\eta }\phi $. Substitution of (\ref{eqn2}) and (\ref{eqn14}) into (\ref{eqn26}) and then integrating both sides yields,
\begin{equation}
\sigma {\rm{ csc(}}\frac{\phi }{\sigma }{\rm{) }}d\phi  = \frac{{{\Lambda ^4}}}{{3{H_i}^2}}\;\frac{{d\eta }}{\eta }.
\label{eqn24}
\end{equation}

Solving for $\phi \left( \eta  \right)$,
\begin{equation}
\phi \left( \eta  \right) = 2\sigma \arctan [{c_2}{\rm{ }}{\eta ^{ - \frac{{{\Lambda ^4}}}{{3{H_i}^2{\sigma ^2}}}}}],
\label{eqn25}
\end{equation}
where ${c_2}$ is the integration constant. By using the model of inflation at which the potential is very small at the end of inflation, ${\eta _{end}}$,
\begin{equation}
V\left( {{\phi _{end}}} \right) = {\Lambda ^4}\left( {1 + \cos (2\arctan [{c_2}{\rm{ }}{\eta _{end}}^{ - \frac{{{\Lambda ^4}}}{{3{H_i}^2{\sigma ^2}}}}])} \right) \ll 1.
\label{eqn26}
\end{equation} 
Hence, the argument of cos should be $\pi$. Thus, the integration constant has to be 
\begin{equation}
{c_2} \gg {\eta _{end}}^{\frac{{{\Lambda ^4}}}{{3{H_i}^2{\sigma ^2}}}}.
\label{eqn27}
\end{equation} 
	
The form (\ref{eqn25}) and the limit (\ref{eqn27}) will be used to derive the coupling function, $f\left( \eta  \right)$, and then calculate the electromagnetic spectra in the de Sitter model approximation, in section \ref{sec:PMF in NI}.

\subsection{NI under a power law expansion}
\label{subsec:sr in PL}

To have a more optimal slow roll analysis that has a smooth exit from inflation, the Hubble parameter can be written as a function of $\phi $, $H(\phi )$. If the field falls below a certain value, it starts to oscillate and then converts to particles in the reheating era, right after inflation. Plugging (\ref{eqn6}) and (\ref{eqn21}) into (\ref{eqn11}), yields
\begin{equation}
a(\eta ) = {l_0}{\left| \eta  \right|^{ - 1 - \frac{1}{{2{\zeta ^2}}}{{\left( {\frac{{\sin (\varpi )}}{{1 + \cos (\varpi )}}} \right)}^2}}},
\label{eqn28}
\end{equation}
where, $\varpi {\rm{ = arc}}\cos \left[ {1 - \left( {\frac{{4{\zeta ^2}}}{{1 + 2{\zeta ^2}}}} \right)\exp ( - N/{\zeta ^2})} \right]$. The Hubble parameter is then
\begin{equation}
H(\eta ) = \frac{{a'(\eta )}}{{{a^2}(\eta )}} \simeq  - \frac{{(1 + \frac{1}{{2{\zeta ^2}}}{{\left( {\frac{{\sin (\varpi )}}{{1 + \cos (\varpi )}}} \right)}^2})}}{{{l_0}}}{\left| \eta  \right|^{\frac{1}{{2{\zeta ^2}}}{{\left( {\frac{{\sin (\varpi )}}{{1 + \cos (\varpi )}}} \right)}^2}}}.
\label{eqn29}
\end{equation}
For $N \ge 50$, and $\zeta  \le 2$, we have $\varpi  \approx 0$. Then we end up with $a(\eta ) \approx {l_0}{\left| \eta  \right|^{ - 1}}$, and $H \approx const$, which is the same as the de Sitter model.
	
However, in the more general solution, one can substitute (\ref{eqn28}) and (\ref{eqn29}) into (\ref{eqn5}) and solve for $\phi $, to yield,
\begin{equation}
\phi \left( \eta  \right)  =  2\sigma \arctan [{c_2} \times {\rm{ exp}}\left\{ { - \frac{{{l_0}^2{{\left( {{{\rm{e}}^{\frac{N}{{{\zeta ^2}}}}} - 2{\zeta ^2} + 2{{\rm{e}}^{\frac{N}{{{\zeta ^2}}}}}{\zeta ^2}} \right)}^2}{\eta ^{ - \frac{2}{{{{\rm{e}}^{\frac{N}{{{\zeta ^2}}}}} - 2{\zeta ^2} + 2{{\rm{e}}^{\frac{N}{{{\zeta ^2}}}}}{\zeta ^2}}}}}{\Lambda ^4}}}{{6\left( {1 + {{\rm{e}}^{\frac{N}{{{\zeta ^2}}}}} - 2{\zeta ^2} + 2{{\rm{e}}^{\frac{N}{{{\zeta ^2}}}}}{\zeta ^2}} \right)\sigma }}} \right\}],
\label{eqn30}
\end{equation}
where, ${c_2}$ is the integration constant.  Using the constraint, $V(\phi ({\eta _{end}})) \ll 1$, implies that,
\begin{equation}
{c_2} \gg {\rm{ exp}}\left\{ {\frac{{{l_0}^2{{\left( {{{\rm{e}}^{\frac{N}{{{\zeta ^2}}}}} - 2{\zeta ^2} + 2{{\rm{e}}^{\frac{N}{{{\zeta ^2}}}}}{\zeta ^2}} \right)}^2}{\eta ^{ - \frac{2}{{{{\rm{e}}^{\frac{N}{{{\zeta ^2}}}}} - 2{\zeta ^2} + 2{{\rm{e}}^{\frac{N}{{{\zeta ^2}}}}}{\zeta ^2}}}}}{\Lambda ^4}}}{{6\left( {1 + {{\rm{e}}^{\frac{N}{{{\zeta ^2}}}}} - 2{\zeta ^2} + 2{{\rm{e}}^{\frac{N}{{{\zeta ^2}}}}}{\zeta ^2}} \right)\sigma }}} \right\}.
\label{eqn31}
\end{equation}
By using (\ref{eqn30}) and (\ref{eqn31}) we can derive the coupling function, $f(\eta )$ , but it is very complicated and actually not needed, as the simple de Sitter approximation is sufficient to investigate PMF under NI.

\section{The PMF generated in natural inflation}
\label{sec:PMF in NI}

Starting from Lagrangian (\ref{eqn1}), the equation of motion for the electromagnetic field, ${A_\mu }$,
\begin{equation}
{\partial _\mu }\left[ {\sqrt { - g} {g^{\mu \nu }}{g^{\alpha \beta }}{f^2}\left( {\phi ,t} \right){F_{\nu \beta }}} \right] = 0,
\label{eqn32}
\end{equation}
where, $g$ is the determinant of the flat FRW space-time metric ${g_{\mu \nu }}$. In conformal time, (\ref{eqn32}) can be written as
\begin{equation}
{{\rm{A''}}_i}\left( {\eta ,x} \right) + 2\frac{{f'}}{f}{{\rm{A'}}_i}\left( {\eta ,x} \right) - {a^2}\left( \eta  \right){\rm{\;}}{\partial _j}{\partial ^j}{{\rm{A}}_i}\left( {\eta ,x} \right) = 0
\label{eqn33}
\end{equation}
Define the function, ${{\rm{\bar A}}_i}\left( {\eta ,x} \right) = f\left( \eta  \right){{\rm{A}}_i}\left( {\eta ,x} \right)$, and using the quantization of ${{\rm{\bar A}}_i}$ in terms of creation and annihilation operators, ${b^\dag }_\lambda $ and ${b_\lambda }\left( k \right)$, as,
\begin{equation}
\label{eqn34}
{{\rm{\bar A}}_i}\left( {\eta ,x} \right) = \int {\frac{{{d^3}k}}{{{{\left( {2\pi } \right)}^{3/2}}}}\mathop \sum \limits_{\lambda  = 1}^2 {\varepsilon _{i\lambda }}\left( k \right)\times
[{b_\lambda }\left( k \right){\cal A}\left( {\eta ,k} \right){e^{ik.x}} + {b^\dag }_\lambda \left( k \right){{\cal A}^*}\left( {\eta ,k} \right){e^{ - ik.x}}]},
\end{equation}
where, ${\varepsilon _{i\lambda }}$ is the transverse polarization vector, and $k = \frac{{2\pi }}{\lambda }$, is the commoving wave number. Hence, Eq.(\ref{eqn33}) can be written as,
\begin{equation}
{\cal A}''\left( {\eta ,k} \right) + \left( {{k^2} - Y\left( \eta  \right)} \right){\cal A}\left( {\eta ,k} \right) = 0,
\label{eqn35}
\end{equation}
where, $Y\left( \eta  \right) = \frac{{f''}}{f}$.

The magnetic and electric spectra can be calculated \cite{7} respectively by, 
\begin{equation}
\frac{{d{\rho _B}}}{{d{\rm{ln}}k}} = \frac{1}{{2{\pi ^2}}}{\left( {\frac{k}{a}} \right)^4}k{\left| {{\cal A}\left( {\eta ,k} \right)} \right|^2}.
\label{eqn36}
\end{equation}
\begin{equation}
\frac{d{\rho_E}}{d{\rm{ln}}k} = \frac{f^2}{2\pi ^2}\frac{{{k^3}}}{{{a^4}}}{\left| {{\left[ \frac{{\cal A}\left( {\eta ,{\rm{ }}k} \right)}{f} \right]}'} \right|^2}.
\label{eqn37}
\end{equation}
Therefore, we need first to define the coupling function, $f\left( \eta  \right)$, in order to solve for the electromagnetic vector field, ${A_\mu }$. We will assume that the relation between the coupling function and scale factor is of the power law form \cite{7}, $f\left( \eta  \right){\rm{\;}} \propto {\rm{\;}}{a^\alpha }$. Then, by combining (\ref{eqn3}) and (\ref{eqn4}) in the slow roll limit,
\begin{equation}
f\left( \phi  \right){\rm{\;}} \propto exp\left[ { - \frac{\alpha }{{3\;{M_{{\rm{Pl}}}}^2}}\mathop \smallint \limits_{}^\phi  \frac{{V\left( \phi  \right)}}{{V'\left( \phi  \right)}}d\phi } \right].
\label{eqn38}
\end{equation}
Substituting (\ref{eqn2}) into (\ref{eqn38}) gives,
\begin{equation}
f\left( {\phi \left( \eta  \right)} \right){\rm{\;}} = D{\rm{ }}\sin {\left[ {\frac{{\phi (\eta )}}{{2\sigma }}} \right]^{\frac{{2\alpha {\sigma ^2}}}{{3{M_{{\rm{Pl}}}}^2}}}}
\label{eqn39}
\end{equation}
where, $D$, is a coupling constant. As we adopt de Sitter or quasi-de Sitter inflationary expansion, the relation between $\alpha$ and $\gamma$ is $\alpha \simeq -\gamma$, where $f\left( \eta  \right) \propto {\eta^\gamma}$.         

Since the power law expansion approaches de Sitter for relatively high $N{\rm{ ( > 50)}}$, as can be seen from (\ref{eqn28}) and (\ref{eqn29}), in this section we will only solve (\ref{eqn35}) in a simple de Sitter model of expansion. However, using (\ref{eqn30}) to find $f\left( \eta  \right)$ explicitly yields a very complicated $Y\left( \eta  \right)$, and an analytical solution of (\ref{eqn35}) cannot be found.        

The de Sitter approximation was used by \cite{10} to investigate PMF. One can investigate PMF under the de Sitter model by substituting  (25) into (39) and  (35). Hence,
\begin{equation}
Y\left( \eta  \right) = \frac{{f''}}{f} = \frac{\Delta {2\alpha {\eta ^{ - 2 + \frac{{2{\Lambda ^4}}}{{3{\sigma ^2}{H_i}^2}}}}{\Lambda ^4}}}{{81{\sigma ^2}{H_i}^4{M_{{\rm{Pl}}}}^4{{\left( {{c_2}^2 + {\eta ^{\frac{{2{\Lambda ^4}}}{{3{\sigma ^2}{H_i}^2}}}}} \right)}^2}}}
\label{eqn40}
\end{equation}
where,
\begin{equation}
\Delta = 3{c_2}^2{M_{{\rm{Pl}}}}^2\left( {3{\sigma ^2}{H_i}^2 - 2{\Lambda ^4}} \right)+ {\sigma ^2}{\eta ^{\frac{{2{\Lambda ^4}}}{{3{\sigma ^2}{H_i}^2}}}}\left( {9{H_i}^2{M_{{\rm{Pl}}}}^2 + 2\alpha {\Lambda ^4}} \right).
\label{eqn41}
\end{equation}
By using the limit (\ref{eqn27}) and the facts that, $(\eta ,{\eta _{end}}) \ll  - 1$, $\sigma  \approx {M_{{\rm{Pl}}}}$, $\Lambda  \approx {M_{{\rm{GUT}}}} \approx {10^{ - 3}}{M_{{\rm{Pl}}}}$\cite{21} and the upper limit of ${H_i} < 3.6 \times {10^{ - 5}}{M_{{\rm{Pl}}}}$ at 95\% CL \cite{22}, then $\frac{{2{\Lambda ^4}}}{{3{\sigma ^2}{H_i}^2}} \ll 1$. Thus, $Y\left( \eta  \right)$ can be written as,
\begin{equation}
Y\left( \eta  \right) \approx \frac{{6\alpha {\eta ^{ - 2}}{\Lambda ^4}\left( {3{\sigma ^2}{H_i}^2 - 2{\Lambda ^4}} \right)}}{{81{\sigma ^2}{H_i}^4{M_{{\rm{Pl}}}}^2{c_2}^2}}
\label{eqn42}
\end{equation} 

Substituting (\ref{eqn42}) into (\ref{eqn35}),
\begin{equation}
{\cal A}''\left( {\eta ,k} \right) + \left( {{k^2} - \frac{{6\alpha {\eta ^{ - 2}}{\Lambda ^4}\left( {3{\sigma ^2}{H_i}^2 - 2{\Lambda ^4}} \right)}}{{81{\sigma ^2}{H_i}^4{M_{{\rm{Pl}}}}^2{c_2}^2}}} \right){\cal A}\left( {\eta ,k} \right) = 0
\label{eqn43}
\end{equation}
Eq.(\ref{eqn43}) is a Bessel differential equation. Hence, ${\cal A}\left( {\eta ,k} \right)$ can be written as
\begin{equation}
{\cal A}\left( {\eta ,k} \right) = {\left( {k\eta } \right)^{1/2}}\left[ {{C_1}\left( k \right)\;{J_\chi }\left( {k\eta } \right) + {C_2}\left( k \right)\;{J_{ - \chi }}\left( {k\eta } \right)} \right]
\label{eqn44}
\end{equation}
where $\chi $ is given by
\begin{equation}
\chi  = \frac{{\sqrt {27 + \frac{{8\alpha \left( {3{\sigma ^2}{H_i}^2{\Lambda ^4} - 2{\Lambda ^8}} \right)}}{{{c_2}^2{\sigma ^2}{H_i}^4{M_{{\rm{Pl}}}}^2}}} }}{{6\sqrt 3 }}.
\label{eqn45}
\end{equation}
In the long wavelength regime, $k \eta \ll 1$ (outside Hubble radius), Eq.(\ref{eqn44}) can be written as
\begin{equation}
{{\cal A}_{k \eta \ll 1}}\left( {\eta ,k} \right) = {\left( k \right)^{ - 1/2}}\left[ {{D_1}\left( \chi  \right)\;{{\left( { - k\eta } \right)}^{\chi +1/2} } + {D_2}\left( \chi  \right){{\left( { - k\eta } \right)}^{1/2 - \chi }}} \right].
\label{eqn45.1}
\end{equation}
The constants, ${D_1}\left( \chi  \right)$ and ${D_2}\left( \chi  \right)$, can be fixed by using the normalization of ${\cal A}\left( {\eta ,k} \right)$ and other limit, ${\cal A}_{k \eta \gg 1} \left( {\eta ,k} \right) \to {{{e^{ - k\eta }}} \mathord{\left/
 {\vphantom {{{e^{ - k\eta }}} {\sqrt {2k} }}} \right.
 \kern-\nulldelimiterspace} {\sqrt {2k} }}$. They are written as
\begin{equation}
{D_1}\left( \chi  \right) = \frac{{\sqrt \pi  }}{{{2^{\chi + 1}}}}\;\frac{{{e^{ - i\pi (\chi + 1/2) /2}}}}{{{\rm{\Gamma }}\left( {\chi  + 1} \right)\cos \left( {\pi (\chi + 1/2) } \right)}}, {\rm{   }}{D_2}\left( \chi  \right) = \frac{{\sqrt \pi  }}{{{2^{1 - \chi }}}}\;\frac{{{e^{ - i\pi \left( {\chi  + 3/2} \right)/2}}}}{{{\rm{\Gamma }}\left( {1 - \chi } \right)\cos \left( {\pi (\chi + 1/2) } \right)}}.
\label{eqn45.2}
\end{equation} 

The magnetic spectra can be obtained by substituting of (\ref{eqn45.1}) into (\ref{eqn36}). It can be written \cite{7} as,
\begin{equation}
\frac{{d{\rho _B}}}{{d{\rm{ln}}k}} = \frac{{\cal F}(n)}{{2{\pi ^2}}}H^4{\left( {\frac{k}{a H}} \right)^{4+2n}},
\label{eqn45.3}
\end{equation}
where, $\gamma= \chi +1/2$, then $n = \gamma$ if $\gamma \leq 1/2$ and $n = 1-\gamma$ for $\gamma \geq 1/2$. The function ${\cal F} (n)$ can be written as,
\begin{equation}
{\cal F}(n) = \frac{\pi}{2^{2n+1}\rm{\Gamma }^2(n + \frac{1}{2})\cos^2 (\pi n)}.
\label{eqn45.4}
\end{equation}
Similarly, the electric field spectrum can be written as,
\begin{equation}
\frac{{d{\rho _E}}}{{d{\rm{ln}}k}} = \frac{{\cal G}(m)}{{2{\pi ^2}}}H^4{\left( {\frac{k}{a H}} \right)^{4+2m}},
\label{eqn45.5}
\end{equation}
where, $m = \gamma+1$ if $\gamma \leq -1/2$ and $m = -\gamma$ for $\gamma \geq -1/2$. The function ${\cal G} (m)$ can be written as,
\begin{equation}
{\cal G}(m) = \frac{\pi}{2^{3m+3}\rm{\Gamma }^2(m + \frac{3}{2})\cos^2 (\pi m)}.
\label{eqn45.6}
\end{equation}

The scale invariant PMF can be achieved if the magnetic spectrum, $\frac{{d{\rho _B}}}{{d{\rm{ln}}k}} = Constant$. Hence, from Eq.(\ref{eqn45.3}), the values of $\gamma = \lbrace-2, 3\rbrace$. The first value is more acceptable in generating the PMF without huge amount of backreaction and without assuming a small scale of inflation ( as in the case of $\gamma=3$) . Also, for $ \alpha=-3$, the reheating period will be to long, such that the reheating temperature falls to few MeV. For $\alpha=2$, the problem of strong coupling exists. One way to solve this problem, is to assume that, the initial coupling function is much less than the coupling function at the end of inflation, $f\left( \eta_0  \right)\ll f\left( \eta_{end}  \right)\thickapprox1$. This assumption in turns, will create a weak coupling between the gauge field and charges at the end of inflation \cite{7}. For these reasons, the solution, $\alpha=2$, is adopted in this paper.     

\begin{figure}[h]
\includegraphics[width=1\textwidth]{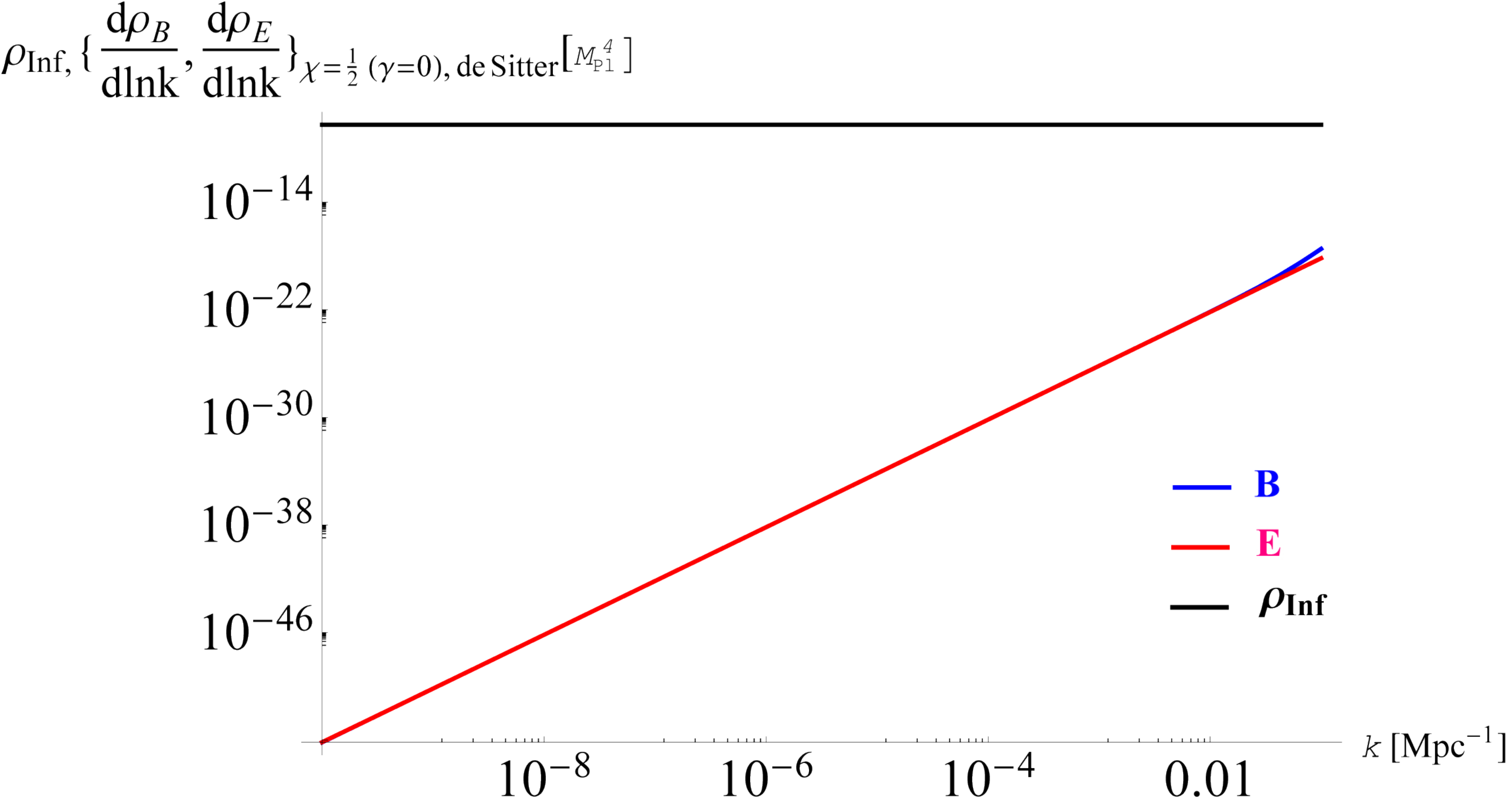}
\caption{ The electromagnetic spectra and the inflationary energy density,  $\rho_{\rm{Inf}}$, in NI at ${\chi} = 1/2$, $\eta {\rm{  = }} - 20, {\rm{ }}\sigma  \approx {M_{{\rm{Pl}}}} = 1$, $,{\rm{ }} \Lambda  = {\rm{1}}{{\rm{0}}^{ - 3}},{\rm{ }}{H_i} = 3.6 \times {\rm{1}}{{\rm{0}}^{ - 5}}{\rm{,  }}\alpha=2$ and $D=1$. The spectrum of electric field is of the same order of magnitude as the spectrum of the magnetic field for $k \eta \ll 1$. At relatively high $k$, they start diverging from each other. However, the energy density of inflation generated by NI, $\rho_{\rm{Inf}}$ is much larger than the electromagnetic energy density.}
\label{f1} 
\end{figure}

Since, $3{\sigma ^2}{H_i}^2{\Lambda ^4} \gg 2{\Lambda ^8}$, then $\chi  = \frac{1}{6}\sqrt {9 + \frac{{8\alpha {\Lambda ^4}}}{{{c_2}^2{H_i}^2{M_{{\rm{Pl}}}}^2}}} $. By using the limit (\ref{eqn27}), we have $\chi  \simeq 1/2$. This value is corresponding to the value of $\gamma = 0$ in \cite{7}. Hence, a scale invariant PMF cannot be generated in NI under the simple de Sitter model of inflation if we impose the limit (\ref{eqn27}). Calculating the electromagnetic spectra shows that, they are almost of the same order of magnitude, at $k \eta \ll 1$, see Fig.\ref{f1}. As, $\gamma=0$ implies that both $n, m=0$. Hence, both magnetic and electric spectra obtained by (\ref{eqn45.3}-\ref{eqn45.5}) will be proportional to $k^{4}$, as can be seen in Fig.\ref{f1}.      

On the other hand, if Eq.(\ref{eqn27}) limit is relaxed and the scale invariance condition is enforced, ${\chi} = 5/2$ $(\gamma = + 3, - 2)$, then ${c_2}$ becomes
\begin{equation}
{c_2} = \sqrt {\frac{1}{{27}}\frac{{\alpha {\rm{ }}{\Lambda ^4}}}{{{H_i}^2{M_{{\rm{Pl}}}}^2}}}
\label{eqn46}
\end{equation}
The coupling function can be written as,
\begin{equation}
f\left( \eta  \right){\rm{\;}} = D{\rm{ }}\sin {\left[ {\arctan [\sqrt {\frac{1}{{27}}\frac{{\alpha {\rm{ }}{\Lambda ^4}}}{{{H_i}^2{M_{{\rm{Pl}}}}^2}}} {\rm{ }}{\eta ^{ - \frac{{{\Lambda ^4}}}{{3{H_i}^2{f^2}}}}}]} \right]^{\frac{{2{f^2}\alpha }}{{3{M_{{\rm{Pl}}}}^2}}}}
\label{eqn47}
\end{equation}

\begin{figure}[tbp]
\includegraphics[width=1\textwidth]{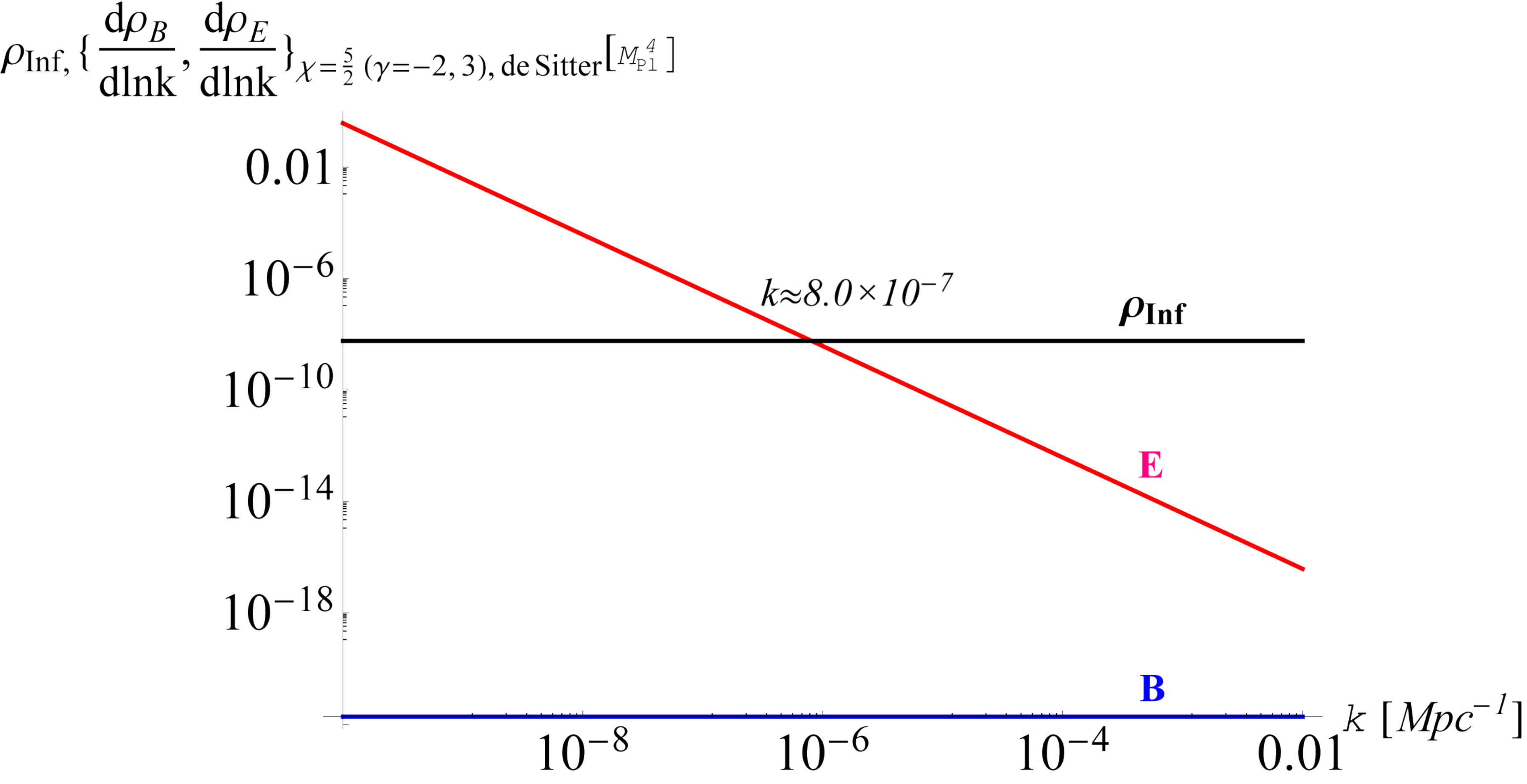}
\caption{The electromagnetic spectra and the inflationary energy density,  $\rho_{\rm{Inf}}$, in NI model at $\chi  = 5/2$, $\eta = - 20$, $\sigma  \approx {M_{{\rm{Pl}}}} = 1$, $\Lambda  = 10^{ - 3}$, ${H_i} = 3.6 \times 10^{ - 5}$, $\alpha =2$ and $D=1$. The spectrum of electric field is much greater than the spectrum of magnetic field for, $k \eta \ll 1$. For $k_{\rm{min}}\gtrsim 8.0\times 10^{-7} \rm{Mpc^{-1}}$ the electromagnetic energy density can go below that of inflation. Hence, the backreaction problem might be avoided for $k_{\rm{min}}<k \ll 1$.}
\label{f2} 
\end{figure}

The electric and magnetic spectra in this case can be seen in Fig.\ref{f2}. We can see that at very long wavelength ($k \eta \ll 1$) the electric field spectrum far exceeds that of the magnetic field and the energy density of the inflation which is generated by NI, $\rho_{\rm{Inf}}$. It may cause the backreaction problem. However, for $k_{\rm{min}}\gtrsim 8.0\times 10^{-7} \rm{Mpc^{-1}}$ the electromagnetic energy can go below that of inflation. Most of the observable scale is above $k_{\rm{min}}$. That range of $k$ includes most of the observable scales according to Planck, 2015. For example, it includes the standard pivot scale, ${k_ * } = 0.05{\rm{Mpc}}^{ - 1}$. Further, it includes some of the cut-off scale, $\ln ({k_c}/{\rm{Mpc}}^{ - 1}) \in [ - 12, - 3]$, chosen by Planck, 2015 \cite{22}. Therefore, the backreaction problem might be avoided in generating PMF by the ${f^2}FF$ model in NI, under de Sitter expansion for $1 \gg k >k_{\rm{min}}$.

\begin{figure}[tbp]
\includegraphics[width=1\textwidth]{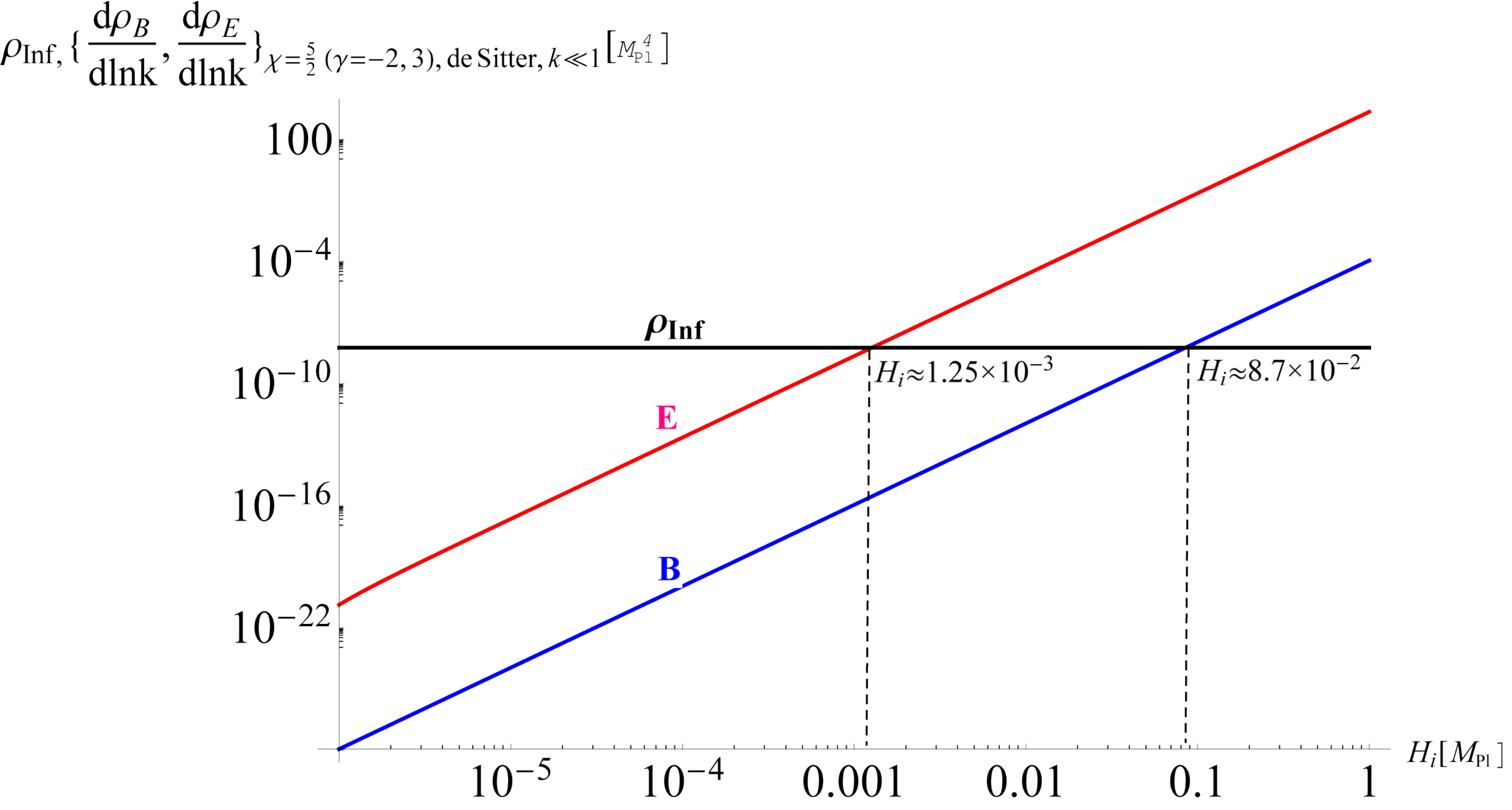}
\caption{The electromagnetic spectra and the inflationary energy density,  $\rho_{\rm{Inf}}$ in NI, as a function of ${H_i}$ at $\chi  = 5/2 $, $\eta  =  - 20$, $\sigma \approx {M_{Pl}} = 1$, $\Lambda  = 10^{ - 3}$, $\alpha = 2$, and $D = 1$. The spectrum of electric field is always much greater than the spectrum of magnetic field and can exceed the energy of inflation for $H_{\rm{min}}\gtrsim 1.25 \times 10^{-3} \rm{M_{\rm{Pl}}}$. This value is well above the upper limit of the Hubble parameter, obtained by Planck, 2015, $H_i \lesssim 3.6 \times 10^{-5} M_{\rm{Pl}}$. Below $H_{\rm{min}}$, all electromagnetic spectra will be less than $\rho_{\rm{Inf}}$. Hence, that might avoid the backreaction problem.}
\label{f3} 
\end{figure}

Likewise, plotting the electromagnetic spectra as a function of the Hubble parameter ${H_i}$, shows that the electric field is always greater than the magnetic field and can exceed the energy of inflation for $H_{\rm{min}}\gtrsim 1.25\times 10^{-3} \rm{M_{\rm{Pl}}}$.(See Fig.\ref{f3}.). This value is well above the upper limit of the Hubble parameter, obtain by Planck, 2015, $H_i\lesssim 3.6 \times 10^{-3} \rm{M_{\rm{Pl}}}$ \cite{22}. Hence, this model can be free from the backreaction problem.   

\begin{figure}[tbp]
\includegraphics[width=1\textwidth]{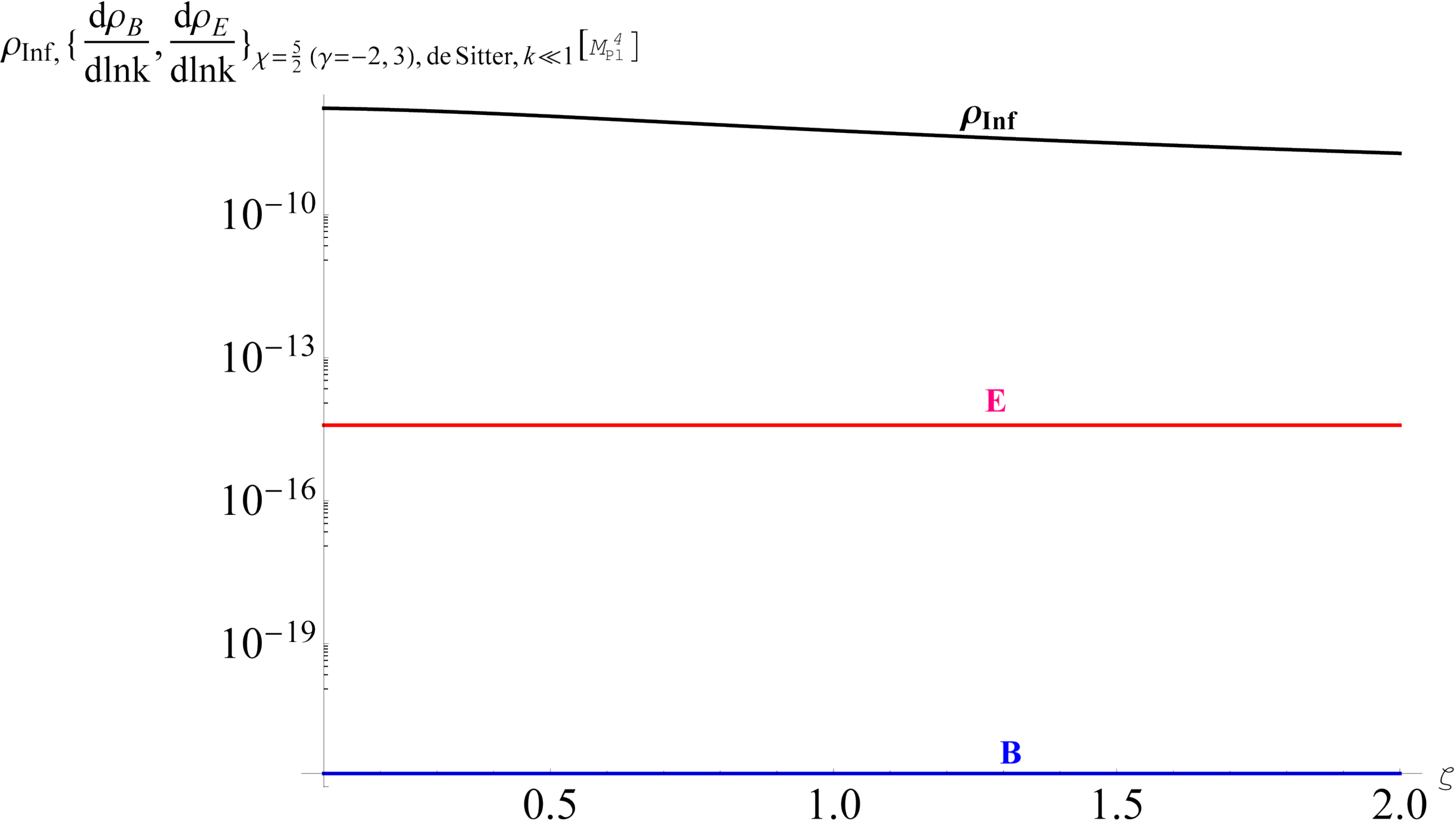}
\caption{The electromagnetic spectra and the inflationary energy density,  $\rho_{\rm{Inf}}$, in NI model as a function of $\zeta  = \sigma /{M_{{\rm{Pl}}}}$ at $\chi  = 5/2$, $\eta  =  - 20$, $\Lambda  = {10^{ - 3}}$, $\alpha = 2$, ${H_i} = 3.6 \times 10^{ - 5}M_{\rm{Pl}}$ and $D = 1$. The spectrum of electric field is always much greater than the spectrum of magnetic field for $k \ll 1$. However, both electric and mgnetic spectra are much less than $\rho_{\rm{Inf}}$. The value $\zeta $ play an effective role in the characteristics of the natural inflation and its implications.}
\label{f4} 
\end{figure}

Similarly, plotting energy density of inflation and the electromagnetic spectra as a function of $\zeta$ clearly shows that the backreaction problem can be avoided for the possible values (See Fig.\ref{f4}.). The value $\zeta $ play an effective role in the characteristics of the natural inflation and its implications. For example, in the case of $\zeta\gg1$, the natural inflation behaves like quadratic inflation \cite{19}.   

\begin{figure}[tbp]
\includegraphics[width=1\textwidth]{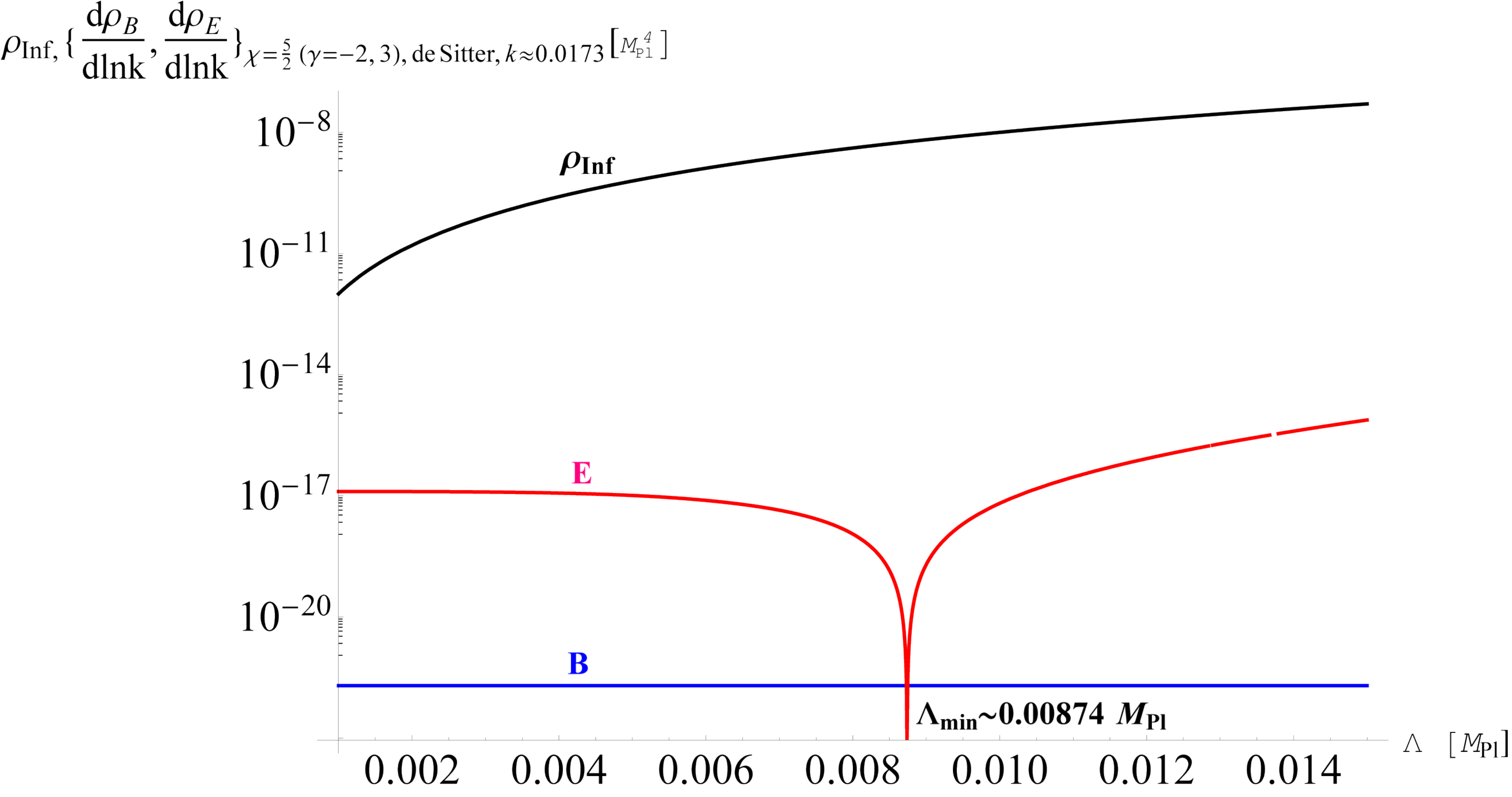}
\caption{The electromagnetic spectra and the inflationary energy density,  $\rho_{\rm{Inf}}$, in NI, as a function of $\Lambda $ at $\chi  = 5/2$, $\eta  =  - 20$, $ \sigma  \approx {M_{{\rm{Pl}}}} = 1$, $\alpha = 2$, $D = 1$, and $k = {10^{ - 3}}{\rm{Mpc}}^{-1}$. The spectrum of electric field falls below the spectrum of magnetic field around $\Lambda  = 0.00874{M_{{\rm{Pl}}}}$. However, both of them are much less than $\rho_{\rm{Inf}}$, which may avoid the backreaction problem.}
\label{f5} 
\end{figure}

Finally, one can analyze the shape of the electromagnetic spectra as a function of $\Lambda$. As seen in Fig.\ref{f5}, there is a narrow range of $\Lambda $ ($\sim 0.00874 M_{\rm{Pl}}$), at which the electric fields can even fall below the magnetic field. 

\begin{figure}[tbp]
\includegraphics[width=1\textwidth]{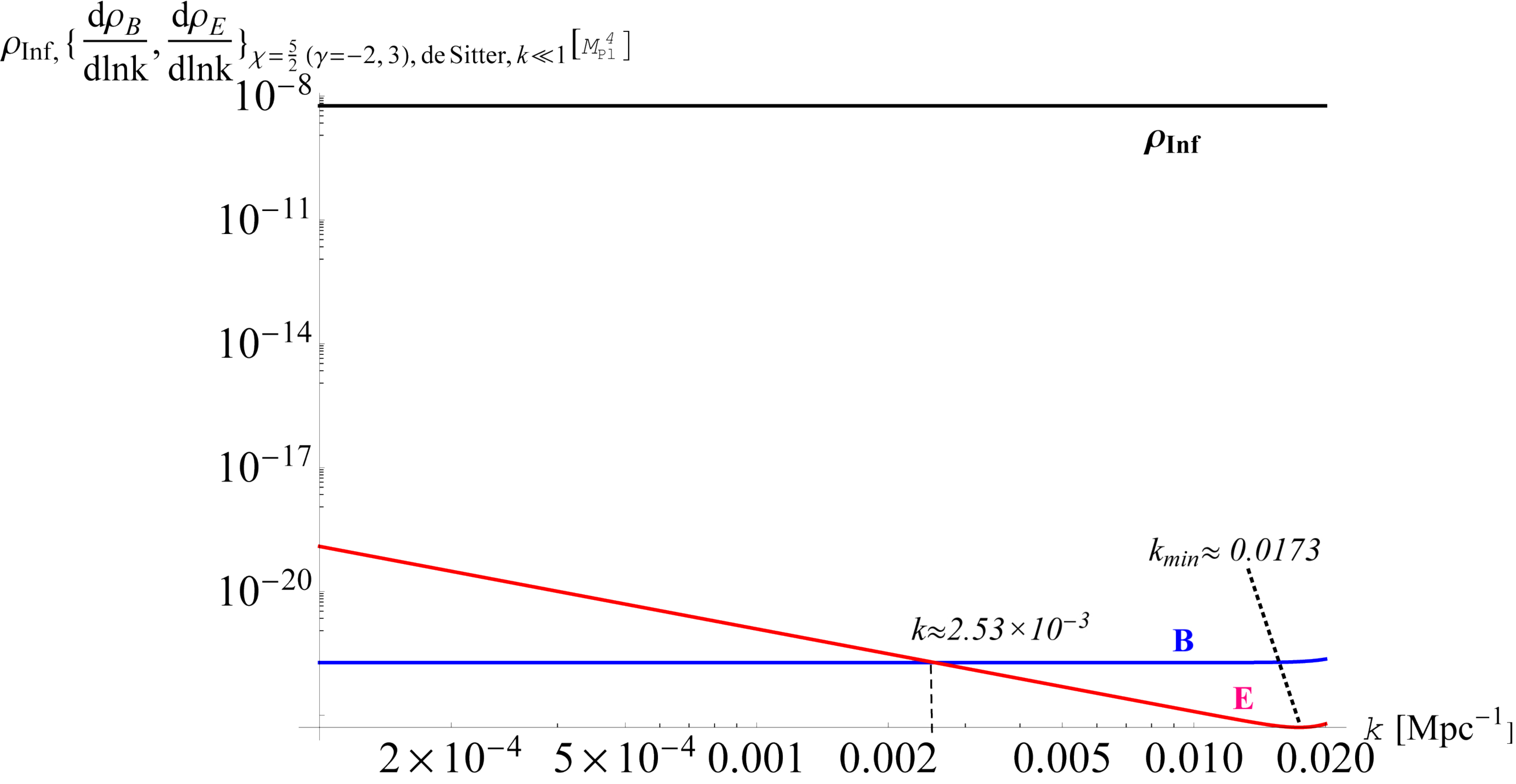}
\caption{The electromagnetic spectra and the inflationary energy density,  $\rho_{\rm{Inf}}$, in NI as a function of $k$ at $\chi  = 5/2$, $\eta  =  - 20$, $\sigma  \approx {M_{{\rm{Pl}}}} = 1$, $\alpha = 2$, $D = 1$, and $\Lambda  = 0.00874{M_{{\rm{Pl}}}}$. The spectrum of electric field falls below the spectrum of magnetic field on the range of, $k \gtrsim 2.53 \times {10^{ - 3}}{\rm{Mpc}}^{ - 1}$, at which the backreaction problem can be avoided.}
\label{f6} 
\end{figure}

In order to decide the range of $k$ for which the electric field energy is less than the magnetic fields, one can plot the electromagnetic spectra as a function of $k$, as in Fig.\ref{f6}. The range is $k \gtrsim 2.53 \times {10^{ - 3}}{\rm{Mpc}}^{ - 1} $, for  $k \eta \ll 1$, around ${k_{\min }}\sim 0.0173{\rm{Mpc}}^{ - 1}$. As we choose ${M_{{\rm{Pl}}}} = 1 {\rm{ (}} \approx {\rm{1}} \times {\rm{1}}{{\rm{0}}^{19}}{\rm{GeV)}}$, the appropriate values of $\Lambda $ is in the order of ${M_{{\rm{GUT}}}} \sim {10^{16}}{\rm{GeV }}$ that fits with the results of Ref.\cite{19}.  

On the other hand, the range of $k$ includes most of the observable scales according to Planck, 2015. For example, it includes the standard pivot scale, ${k_ * } = 0.05{\rm{Mp}}{{\rm{c}}^{ - 1}}$. Further, it includes some of the cut-off scale, $\ln ({k_c}/{\rm{Mp}}{{\rm{c}}^{ - 1}}) \in [ - 12, - 3]$, chosen by Planck, 2015 \cite{22}. However, the relatively narrow range of $k$ at which this situation is valid, may cause serious challenge to this model. This is so because after sufficient number of e-foldings the wave number may go below $k < {10^{-3}}{\rm{Mpc}}^{ - 1}$. 

\section{Summary and conclusions}
\label{sec:Summary}

PMF can be generated by the simple inflation model ${f^2}FF$, Eq.(\ref{eqn1}), in the standard models of inflation, and requires the breaking of the conformal symmetry of the electromagnetism. In this paper, we used the same method used in \cite{7} to investigate the PMF in natural inflation. We first presented the slow roll analysis of the NI. Unlike the PMF on large field inflation (LFI) \cite{23} and similar to PMF on ${R^2}$-inflation \cite{24}, for sufficiently large number of e-foldings, the power law inflation can lead to the same results as the simple de Sitter model of expansion in NI.  

We find that PMFs can in principle be generated in the NI model for all values of $\zeta  = \sigma /{M_{{\rm{Pl}}}}$. Under $V(0)\thickapprox0$ model of inflation, the scale invariant PMF is unlikely to be generated in the context of NI. That is similar to the case in the context of LFI \cite{23}. However, if this constraint is relaxed, a scale invariant PMF can be achieved in NI. In this case, the magnitude of the PMF spectrum, at $k \eta \ll 1$, is much smaller than the spectrum of the associated electric field. Changing the values of $\eta$ ,$\sigma$ ,$H_i$, $\alpha$, and $D$ does not change this relation. In comparison with the inflationary energy density,  $\rho_{\rm{Inf}}$, in NI and the upper bound of the energy density of inflation derived from WMAP7, ${\left( {{\rho _{{\rm{end}}}}} \right)_{{\rm{CMB}}}} < 2.789 \times {10^{ - 10}}M_{{\rm{Pl}}}^4$ \cite{27}, the energy of the electric field may exceed the energy scale of inflation at $k \lesssim 8.0\times 10^{-7} \rm{Mpc^{-1}}$ and $H_i \gtrsim 1.25\times 10^{-3} \rm{M_{\rm{Pl}}}$. That may prevent inflation from occurring at all. This is the problem of backreaction. One can conclude that for small enough value of $k$, this problem cannot be avoided in the ${f^2}FF$ model under natural inflation.
 
On the other hand for $k > 8.0\times 10^{-7} \rm{Mpc^{-1}}$ and $H_i \lesssim 1.25\times 10^{-3} \rm{M_{\rm{Pl}}}$ , both electric and magnetic energy densities can fall below the inflationary energy density. In this case, one can consider these values as, respectively, a lower bound of $k$ and an upper bound of $H_i$ for a backreaction-free model of PMF.  Moreover, these scales include most of the observable ranges of $k$ and $H_i$. 

Furthermore, there is a range of  $\Lambda_{\rm{min}} (\sim 0.00874{M_{\rm{Pl}}})$, and $ k\gtrsim 2.53 \times 10^{ - 3}{\rm{Mpc}}^{- 1}$, at which the energy density of the electric field can even fall below the energy density of the magnetic field. Again these values lie on the observable range of $k$ and the anticipated scale of $Lambda$. Therefore, the problem of backreaction can be avoided in these ranges of values. However, the relatively short range of $k$, presents a serious challenge to the viability of this model. One way to extend this research is to include the effect of reheating era and then to calculate the present value of PMF generated in NI as we do in the context of ${R^2}$-inflation \cite{24}.

\section*{Acknowledgments}
We would like to thank Bharat Ratra for useful comments. This work is supported in part by the Department of Physics and Astronomy in The University of Texas at San Antonio.

\end {document}